\documentclass[12pt]{article}

\usepackage{amsmath,amssymb,amscd,amsbsy,amsgen,amsopn,amstext,
amsxtra}

\usepackage[dvips]{graphicx}

\begin{document}

\begin{flushright}
\end{flushright}

\begin{center}
{\Large{\bf Geometric phase and chiral anomaly; their basic differences\footnote{Invited talk given at the Conference in 
Honor of CN Yang's 85th Birthday, October 31-November 3, 2007, Nanyang Technological University, Singapore }}}
\end{center}
\vskip .5 truecm
\centerline{\bf  Kazuo Fujikawa }
\vskip .4 truecm
\centerline {\it Institute of Quantum Science, College of 
Science and Technology}
\centerline {\it Nihon University, Chiyoda-ku, Tokyo 101-8308, 
Japan}
\vskip 0.5 truecm

\makeatletter
\@addtoreset{equation}{section}
\def\theequation{\thesection.\arabic{equation}}
\makeatother

\begin{abstract}
All the geometric phases are shown to be topologically trivial by
using the second quantized formulation. The exact hidden local 
symmetry in the Schr\"{o}dinger equation, which was hitherto 
unrecognized, controls the holonomy associated with both of the 
adiabatic and non-adiabatic geometric phases.
The second quantized formulation is located in between
the first quantized formulation and the field theory, and thus 
it is convenient to compare the geometric phase with the chiral 
anomaly in field theory. It is shown that these two notions are 
completely different.   
\end{abstract}

\section{Introduction}

Phases are intriguing notions, as was emphasized by C.N. Yang on various
occasions. Here we discuss two phases, and the first phase is 
the geometric phase in quantum 
mechanics~\cite{higgins, mead, berry, simon, wilczek,
aharonov, samuel,sjoqvist,singh} for which we present the 
recent developments on the basis of the second quantized 
formulation of all the geometric phases~\cite{fuji-deguchi, 
fujikawa2,fujikawa3,fujikawa4}. The second phase is the chiral 
anomaly in field 
theory~\cite{bell, adler, wess, fujikawa}, which is by now well
understood~\cite{fujikawa-suzuki}. The second quantized 
formulation is located in between
the first quantized formulation and the field theory, and thus 
it is convenient to compare the geometric phase with the chiral 
anomaly in field theory~\cite{nelson, stone, fujikawa5}. 

We then show\\
1. A unified treatment of adiabatic and non-adiabatic geometric 
phases is possible in the second quantized formulation by using 
the exact hidden local ({\em i.e.},time-dependent) symmetry in 
the Schr\"{o}dinger equation.\\
2. The topology of all the geometric phases is trivial
by using an exactly solvable example.\\
3. Geometric phases in the Schr\"{o}dinger problem and the 
chiral anomaly in field theory are completely different.

\section{Second quantized formulation}

We start with defining an {\em arbitrary} complete basis set
\begin{eqnarray}
\int d^{3}x v_{n}^{\star}(t,\vec{x})v_{m}(t,\vec{x})=
\delta_{nm}
\end{eqnarray}
and expand the field operator $\hat{\psi}(t,\vec{x})$ as
\begin{eqnarray}
\hat{\psi}(t,\vec{x})=\sum_{n}\hat{b}_{n}(t)v(t,\vec{x}).
\end{eqnarray}
The action
\begin{eqnarray}
S=\int_{0}^{T}dtd^{3}x[
\hat{\psi}^{\star}(t,\vec{x})i\hbar\frac{\partial}{\partial t}
\hat{\psi}(t,\vec{x})-\hat{\psi}^{\star}(t,\vec{x})
\hat{H}(t)\hat{\psi}(t,\vec{x})]
\end{eqnarray}
which gives rise to the field equation 
\begin{eqnarray}
i\hbar\frac{\partial}{\partial t}\hat{\psi}(t,\vec{x})=
\hat{H}(t)\hat{\psi}(t,\vec{x})
\end{eqnarray}
then becomes 
\begin{eqnarray}
S=\int_{0}^{T}dt\{\sum_{n}
\hat{b}^{\dagger}_{n}(t)i\hbar\partial_{t}\hat{b}_{n}(t)
-\hat{H}_{eff} \}.
\end{eqnarray}
The effective Hamiltonian is given by 
\begin{eqnarray}
\hat{H}_{eff}(t)&=&\sum_{n,m}\hat{b}_{n}^{\dagger}(t)[
\int d^{3}x v_{n}^{\star}(t,\vec{x})\hat{H}(t)v_{m}(t,\vec{x})
\nonumber\\
&&-\int d^{3}x v_{n}^{\star}(t,\vec{x})
i\hbar\frac{\partial}{\partial t}v_{m}(t,\vec{x})]
\hat{b}_{m}(t)
\end{eqnarray}
and the canonical commutation relations
$[\hat{b}_{n}(t), \hat{b}^{\dagger}_{m}(t)]_{\mp}
=\delta_{n,m}$, but statistics (fermions or bosons)
is not important in our application.

The Schr\"{o}dinger picture $\hat{{\cal H}}_{eff}(t)$ is 
obtained by replacing $\hat{b}_{n}(t)$ with 
$\hat{b}_{n}(0)$ in $\hat{H}_{eff}(t)$ (2.6).
Then the evolution operator is given by\cite{fujikawa2}  
\begin{eqnarray}
&&\langle m|T^{\star}\exp\{-\frac{i}{\hbar}\int_{0}^{t}
\hat{{\cal H}}_{eff}(t)
dt\}|n\rangle\nonumber\\ 
&&=
\langle m(t)|T^{\star}\exp\{-\frac{i}{\hbar}\int_{0}^{t}
\hat{H}(\hat{\vec{p}}, \hat{\vec{x}},  
X(t))dt \}|n(0)\rangle 
\end{eqnarray}
with time ordering symbol $T^{\star}$.
In the second quantized formulation on the left-hand side we 
have $|n\rangle=\hat{b}_{n}^{\dagger}(0)|0\rangle$, and in 
the first quantized formulation on the right-hand side we have
$\langle\vec{x}|n(t)\rangle=v_{n}(t,\vec{x})$.

The exact Schr\"{o}dinger probability amplitude which satisfies
  $i\hbar\partial_{t}
\psi_{n}(t,\vec{x})=\hat{H}(t)\psi_{n}(t,\vec{x})$ with
$\psi_{n}(0,\vec{x})=v_{n}(0,\vec{x})$ is given by
\begin{eqnarray}
\psi_{n}(t,\vec{x})&=&\langle0|\hat{\psi}(t,\vec{x})
\hat{b}^{\dagger}_{n}(0)|0\rangle\nonumber\\
&=&\sum_{m} v_{m}(t,\vec{x})
\langle 0|\hat{b}_{m}(t)\hat{b}^{\dagger}_{n}(0)|0\rangle\nonumber\\
&=&\sum_{m} v_{m}(t,\vec{x})
\langle m|T^{\star}\exp\{-\frac{i}{\hbar}\int_{0}^{t}
\hat{{\cal H}}_{eff}(t)dt\}|n\rangle
\end{eqnarray}
which is confirmed by using the relation
$i\hbar\partial_{t}
\hat{\psi}(t,\vec{x})=\hat{H}(t)\hat{\psi}(t,\vec{x})$ in (2.4).
We note that the general geometric terms automatically appear 
as the second terms in the {\em exact} $\hat{{\cal H}}_{eff}(t)$
in (2.8). See $\hat{H}_{eff}(t)$ in (2.6).

\subsection{ Hidden local symmetry}

Since the basic field variable is written as 
$\hat{\psi}(t,\vec{x})
=\sum_{n}\hat{b}_{n}(t)v_{n}(t,\vec{x})$, we have an exact 
{\em hidden} local (i.e., time dependent) 
symmetry~\cite{fujikawa2}
\begin{eqnarray}
&&v_{n}(t,\vec{x})\rightarrow v^{\prime}_{n}(t, \vec{x})=
e^{i\alpha_{n}(t)}v_{n}(t,\vec{x}),\nonumber\\
&&\hat{b}_{n}(t) \rightarrow \hat{b}^{\prime}_{n}(t)=
e^{-i\alpha_{n}(t)}\hat{b}_{n}(t), \ \ \ \ n=1,2,3,... 
\end{eqnarray}
which keeps $\hat{\psi}(t,\vec{x})$ invariant.
This symmetry means arbitrariness in the choice of the coordinates in the functional space. The
Schr\"{o}dinger amplitude $\psi_{n}(t,\vec{x})
=\langle0|\hat{\psi}(t,\vec{x})\hat{b}^{\dagger}_{n}(0)|0\rangle$ is 
then transformed as
\begin{eqnarray} 
\psi^{\prime}_{n}(t,\vec{x})=e^{i\alpha_{n}(0)}
\psi_{n}(t,\vec{x})
\end{eqnarray}
under the hidden symmetry for any $t$. Namely, it gives the ray 
representation with a constant phase.
We thus have the enormous hidden local symmetry behind the ray 
representation, which was not recognized in the past. 
The product 
$\psi_{n}(0,\vec{x})^{\star}\psi_{n}(T,\vec{x})$ is then
manifestly gauge invariant for a periodic system.

If one chooses a specific basis
\begin{eqnarray} 
\hat{H}(X(t))v(\vec{x};X(t))={\cal E}_{n}(X(t))v(\vec{x};X(t))
\end{eqnarray}
in (2.1) for a periodic Hamiltonian 
$\hat{H}(X(0))=\hat{H}(X(T))$ and assumes "diagonal dominance" 
in the effective Hamiltonian, we have from (2.8)
\begin{eqnarray}
\psi_{n}(t,\vec{x})
\simeq v_{n}(\vec{x};X(t))
\exp\{-\frac{i}{\hbar}\int_{0}^{t}[{\cal E}_{n}(X(t))
-\langle n|i\hbar\frac{\partial}{\partial t}|n\rangle]dt\}
\end{eqnarray}
which reproduces the result of the conventional adiabatic 
approximation.

This shows that 
\begin{eqnarray}
{\rm Adiabatic\ approximation} = {\rm Approximate\ diagonalization
\ of} \  H_{eff}\nonumber
\end{eqnarray}
and thus the geometric phases are {\em dynamical}, i.e., a part 
of the Hamiltonian. In fact, 
it has been recently shown that the second quantized formulation
nicely resolves some of the subtle problems in the conventional
adiabatic approximation~\cite{fujikawa6}.

In the  adiabatic approximation (2.12), we have a gauge invariant 
quantity (for a general choice of the hidden local symmetry) 
\begin{eqnarray}
\psi_{n}(0,\vec{x})^{\star}\psi_{n}(T,\vec{x})
&=&v_{n}(0,\vec{x}; X(0))^{\star}v_{n}(T,\vec{x};X(T))
\nonumber\\
&&\times\exp\{-\frac{i}{\hbar}\int_{0}^{T}[{\cal E}_{n}(X(t))
-\langle n|i\hbar\frac{\partial}{\partial t}|n\rangle]dt\}.
\end{eqnarray}
If one chooses a specific hidden local gauge such that 
 $v_{n}(T,\vec{x};X(T))
=\\ v_{n}(0,\vec{x}; X(0))$, the pre-factor 
$v_{n}(0,\vec{x}; X(0))^{\star}v_{n}(T,\vec{x};X(T))$ becomes 
real and positive and thus the factor on the exponential in 
(2.13) represents the entire gauge invariant phase. This unique
gauge invariant quantity reproduces the conventional adiabatic phase\cite{berry,simon}. 

\subsection{ Parallel transport and holonomy}

The parallel transport of $v_{n}(t,\vec{x})$ is defined by
\begin{eqnarray}
\int d^{3}x v^{\dagger}_{n}(t,\vec{x})\frac{\partial}{\partial t}
v_{n}(t,\vec{x})=0
\end{eqnarray}
which is derived from the conditions 
\begin{eqnarray}
\int d^{3}x v^{\dagger}_{n}(t,\vec{x})
v_{n}(t+\delta t,\vec{x})={\rm real\ and\ positive}
\end{eqnarray} 
and
\begin{eqnarray}
\int d^{3}x v^{\dagger}_{n}(t+\delta t,\vec{x})
v_{n}(t+\delta t,\vec{x})
=\int d^{3}x v^{\dagger}_{n}(t,\vec{x})
v_{n}(t,\vec{x}).
\end{eqnarray}
By using the hidden local gauge
$\bar{v}_{n}(t,\vec{x})=e^{i\alpha_{n}(t)}v_{n}(t,\vec{x})$
for a general $v_{n}(t,\vec{x})$, which may not satisfy the 
condition (2.14), the parallel transport condition
\begin{eqnarray}
\int d^{3}x \bar{v}^{\dagger}_{n}(t,\vec{x})\frac{\partial}
{\partial t}
\bar{v}_{n}(t,\vec{x})=0
\end{eqnarray}
gives
\begin{eqnarray}
\bar{v}_{n}(t,\vec{x})
=\exp[i\int_{0}^{t}dt^{\prime}
\int d^{3}x v^{\dagger}_{n}(t^{\prime},\vec{x})
i\partial_{t^{\prime}}v_{n}(t^{\prime},\vec{x})]v_{n}(t,\vec{x}).
\end{eqnarray}
Since $\bar{v}_{n}(t,\vec{x})$ satisfies the parallel transport 
condition, the {\bf holonomy}, i.e., the phase change after 
one cycle, is given by~\cite{fujikawa4}  
\begin{eqnarray}
&&\bar{v}^{\dagger}_{n}(0,\vec{x})
\bar{v}_{n}(T,\vec{x})\nonumber\\
&&=v^{\dagger}_{n}(0,\vec{x})v_{n}(T,\vec{x})
\exp[i\int_{0}^{T}dt^{\prime}
\int d^{3}x v^{\dagger}_{n}(t^{\prime},\vec{x})
i\partial_{t^{\prime}}v_{n}(t^{\prime},\vec{x})].
\end{eqnarray}
This holonomy of {\em basis vectors}, not of the 
Schr\"{o}dinger amplitude, associated with the hidden local 
symmetry determines {\em all} the geometric 
phases. In fact, the adiabatic phase in (2.13) is an example.

\subsection{ Non-adiabatic phase: Cyclic evolution}

The cyclic evolution is defined by~\cite{aharonov} 
\begin{eqnarray} 
&&\int d^{3}x \psi^{\dagger}(t,\vec{x})\psi(t,\vec{x})=1,
\nonumber\\
&&\psi(t,\vec{x})=e^{i\phi(t)}\tilde{\psi}(t,\vec{x}),\ \ \
\tilde{\psi}(T,\vec{x})=\tilde{\psi}(0,\vec{x}).
\end{eqnarray}
namely, $\psi(T,\vec{x})=e^{i\phi}\psi(0,\vec{x})$ with 
$\phi(T)=\phi, \ \ \phi(0)=0$. 

If one chooses the first element of the arbitrary basis set 
$\{v_{n}(t,\vec{x})\}$ in (2.1) such that
$v_{1}(t,\vec{x})=\tilde{\psi}(t,\vec{x})$, one can confirm that
 the exact Schr\"{o}dinger amplitude (2.8) is written as  
\begin{eqnarray}
\psi(t,\vec{x})
&=&v_{1}(t,\vec{x})\exp\{-\frac{i}{\hbar}
[\int_{0}^{t}dt\int d^{3}x v^{\star}_{1}(t,\vec{x})
\hat{H}v_{1}(t,\vec{x})\nonumber\\
&&\hspace{1.5 cm}-\int_{0}^{t}dt\int d^{3}x 
v^{\star}_{1}(t,\vec{x})i\hbar\partial_{t}v_{1}(t,\vec{x})]\}.
\end{eqnarray}
Under the hidden local symmetry of basis vectors, we have
\begin{eqnarray}
\psi(t,\vec{x})\rightarrow e^{i\alpha_{1}(0)}\psi(t,\vec{x})
\end{eqnarray}
and the gauge invariant quantity is given by 
\begin{eqnarray}
&&\psi^{\dagger}(0,\vec{x})\psi(T,\vec{x})\nonumber\\
&&=v^{\star}_{1}(0,\vec{x})v_{1}(T,\vec{x})
\exp\{-\frac{i}{\hbar}
\int_{0}^{T}dt\int d^{3}x[ v^{\star}_{1}(t,\vec{x})
\hat{H}v_{1}(t,\vec{x}) \nonumber\\
&&\hspace{5 cm} -v^{\star}_{1}(t,\vec{x})i\hbar\partial_{t}
v_{1}(t,\vec{x})]\}.
\end{eqnarray}
If one chooses the specific hidden local symmetry 
$v_{1}(0,\vec{x})=v_{1}(T,\vec{x})$,\\
$v^{\star}_{1}(0,\vec{x})v_{1}(T,\vec{x})$ becomes real and 
positive, and the factor
\begin{eqnarray}
\beta=\oint dt \int d^{3}x v^{\star}_{1}(t,\vec{x})
i\frac{\partial}{\partial t}v_{1}(t,\vec{x})
\end{eqnarray}
gives the unique {\em non-adiabatic phase}~\cite{aharonov}.
Eq.(2.23) gives another example of the holonomy (2.19), 
namely, the holonomy of the basis 
vector, not of the Schr\"{o}dinger amplitude, determines the 
non-adiabatic phase in our formulation~\cite{fujikawa3}.

Note that the so-called "projective Hilbert space" and the 
transformation of the Schr\"{o}dinger amplitude~\cite{aharonov}
\begin{eqnarray}
\psi(t,\vec{x})\rightarrow e^{i\omega(t)}\psi(t,\vec{x}),
\end{eqnarray}
which is  not the symmetry of the Schr\"{o}dinger equation, is 
not used in our formulation.  We note that the 
consistency of the ``projective Hilbert space'' (2.25) with the 
superposition principle is not obvious~\cite{fujikawa3}. More about this will be discussed later. 

\subsection{ Non-adiabatic phase: Non-cyclic evolution}

{\it Any} exact Schr\"{o}dinger amplitude is written in the form
\begin{eqnarray}
\psi_{k}(\vec{x},t)
&&=v_{k}(\vec{x},t)\exp\{-\frac{i}{\hbar}\int_{0}^{t}
\int d^{3}x[v^{\dagger}_{k}(\vec{x},t)
\hat{H}(t)v_{k}(\vec{x},t)\nonumber\\
&&\hspace{3 cm}-v^{\dagger}_{k}(\vec{x},t)
i\hbar\frac{\partial}{\partial t}v_{k}(\vec{x},t)]\}
\end{eqnarray}
if one chooses $\{v_{k}(\vec{x},t)\}$ suitably~\cite{fujikawa4}. 
Note, however, the periodicity
\begin{eqnarray}
v_{k}(T,\vec{x})=v_{k}(0,\vec{x})
\end{eqnarray}
is lost in general, and thus {\em non-cyclic}.

In this case, the quantity
\begin{eqnarray}
&&\int d^{3}x
\psi_{k}^{\dagger}(0,\vec{x})\psi_{k}(T,\vec{x})
=\int d^{3}x
v_{k}^{\dagger}(0,\vec{x})v_{k}(T,\vec{x})\nonumber\\
&&\times\exp\{\frac{-i}{\hbar}\int_{0}^{T}dtd^{3}x[
v_{k}^{\dagger}(t,\vec{x})\hat{H}(t)v_{k}(t,\vec{x})
\nonumber\\
&&\hspace{3 cm}
-v_{k}^{\dagger}(t,\vec{x})i\hbar\partial_{t}
v_{k}(t,\vec{x})]\}
\end{eqnarray}
is manifestly invariant under the hidden local symmetry (2.9).
By choosing a suitable hidden symmetry 
$v_{k}(t,\vec{x})\rightarrow
e^{i\alpha_{k}(t)}v_{k}(t,\vec{x})$, one can make the pre-factor
\begin{eqnarray}
\int d^{3}x
v_{k}^{\dagger}(0,\vec{x})v_{k}(T,\vec{x})
\end{eqnarray}
real and positive. It is important that we can make only the 
integrated pre-factor (2.29) real and positive in the present 
non-cyclic case, since one cannot make 
$v_{k}^{\dagger}(0,\vec{x})v_{k}(T,\vec{x})$ real and positive 
by a time dependent gauge transformation for all $\vec{x}$
for the non-cyclic case~\cite{fujikawa2}.
Then the exponential factor in (2.28) defines the unique 
non-cyclic and non-adiabatic phase~\cite{samuel}. We have a 
structure similar to (2.19) in the present non-cyclic case also,
though it may not be called holonomy in a rigorous sense. We 
emphasize that we do not use the projective Hilbert space 
defined by  (2.25) in the present formulation of non-adiabatic 
and non-cyclic geometric phase~\cite{fujikawa4}.

\subsection{Geometric phase for mixed states}

We start with a given hermitian Hamiltonian $\hat{H}(t)$ and 
given ${\cal U}(t)=\\ T^{\star}\exp[-\frac{i}{\hbar}
\int_{0}^{t} \hat{H}(t)dt]$. We employ a diagonal form of the 
density matrix
\begin{eqnarray}
\rho(0)=\sum_{k}\omega_{k}
\psi_{k}(0,\vec{x})\psi^{\dagger}_{k}(0,\vec{x}),
\end{eqnarray}
where the exact Schr\"{o}dinger amplitudes are defined by
\begin{eqnarray}
\psi_{k}(t,\vec{x})=\langle\vec{x}|{\cal U}(t)|k\rangle
=\int d^{3}y\langle\vec{x}|{\cal U}(t)|\vec{y}\rangle 
v_{k}(0,\vec{y}).
\end{eqnarray}
We define the total phases for pure states 
$\psi_{k}(t,\vec{x})$ by
\begin{eqnarray}
\phi_{k}(t)={\rm arg}\int d^{3}x
\psi^{\dagger}_{k}(0,\vec{x})\psi_{k}(t,\vec{x})
\end{eqnarray}
and the complete set of basis vectors in (2.1) by 
\begin{eqnarray}
v_{k}(t,\vec{x})=e^{-i\phi_{k}(t)}\psi_{k}(t,\vec{x})
, \ \  
\int d^{3}xv^{\dagger}_{k}(t,\vec{x})v_{l}(t,\vec{x})
=\delta_{k,l}.
\end{eqnarray}
One can then confirm that the exact Schr\"{o}dinger amplitudes 
are written as
\begin{eqnarray}
\psi_{k}(\vec{x},t)
&=&v_{k}(\vec{x},t)\\
&&\times
\exp\{-\frac{i}{\hbar}\int_{0}^{t}
[\int d^{3}x v^{\dagger}_{k}(\vec{x},t)
\hat{H}(t)v_{k}(\vec{x},t)
-\langle k|i\hbar\frac{\partial}{\partial t}|k\rangle]\}
\nonumber
\end{eqnarray}
with 
\begin{eqnarray}
\langle k|i\hbar\frac{\partial}{\partial t}|k\rangle\equiv
\int d^{3}x v^{\dagger}_{k}(\vec{x},t)
i\hbar\frac{\partial}{\partial t}v_{k}(\vec{x},t).
\end{eqnarray}
 The Schr\"{o}dinger
amplitude $\psi_{k}(t,\vec{x})$ is transformed under 
the hidden local symmetry as 
$\psi_{k}(t,\vec{x})\rightarrow e^{i\alpha_{k}(0)}
\psi_{k}(t,\vec{x})$
independently of $t$ and thus the Schr\"{o}dinger equation is 
invariant under the hidden local symmetry.

The quantity ${\rm Tr}{\cal U}(T)\rho(0)$ is
 then  written as 
\begin{eqnarray}
{\rm Tr}{\cal U}(T)\rho(0)&=&\sum_{k}\omega_{k}
\psi_{k}^{\dagger}(0,\vec{x})\psi_{k}(T,\vec{x})\nonumber\\
&=&\sum_{k}\omega_{k}
v_{k}^{\dagger}(0,\vec{x})v_{k}(T,\vec{x})
\exp\{\frac{i}{\hbar}\int_{0}^{T}dtd^{3}x[
v_{k}^{\dagger}(t,\vec{x})i\hbar\partial_{t}v_{k}(t,\vec{x})
\nonumber\\
&&-v_{k}^{\dagger}(t,\vec{x})\hat{H}(t)v_{k}(t,\vec{x})]\}
\end{eqnarray}
without integration over $\vec{x}$.
If all the pure states perform cyclic evolution with the same 
period $T$, one can choose the hidden local gauge such that
\begin{eqnarray}
v_{k}^{\dagger}(0,\vec{x})v_{k}(T,\vec{x})={\rm real\ and\
positive}
\end{eqnarray}
for all $k$, and the exponential factor in (2.36) exhibits 
the entire geometrical phase together with the 
``dynamical phase'' 
$(1/\hbar)\int_{0}^{T}dtd^{3}xv_{k}^{\dagger}(t,\vec{x})
\hat{H}(t)v_{k}(t,\vec{x})$ of each pure state. 
In practice, the cyclic evolution of all the pure states 
$\psi_{k}(t)$ with a period $T$ may be rather exceptional. For a
generic case, we need to define the phase for  
non-cyclic evolution~\cite{samuel} as the phase of (see (2.28))
\begin{eqnarray}
{\rm Tr}{\cal U}(T)\rho(0)&
=&\sum_{k}\omega_{k}\int d^{3}x
\psi_{k}^{\dagger}(0,\vec{x})\psi_{k}(T,\vec{x})\\
&=&\sum_{k}\omega_{k}\int d^{3}x
v_{k}^{\dagger}(0,\vec{x})v_{k}(T,\vec{x})\nonumber\\
&&\times\exp\{\frac{i}{\hbar}\int_{0}^{T}dtd^{3}x[
v_{k}^{\dagger}(t,\vec{x})i\hbar\partial_{t}v_{k}(t,\vec{x})
-v_{k}^{\dagger}(t,\vec{x})\hat{H}(t)v_{k}(t,\vec{x})]\}
\nonumber
\end{eqnarray}
These quantities (2.36) and (2.38) are manifestly invariant 
under the hidden local symmetry~\cite{fujikawa4}, and thus not only the total 
phase ${\rm arg Tr}{\cal U}(T)\rho(0)$ but also the visibility 
$|{\rm Tr}{\cal U}(T)\rho(0)|$ in the interference 
pattern~\cite{sjoqvist} 
\begin{eqnarray}
I\propto 1+ |{\rm Tr}{\cal U}(T)\rho(0)|\cos[\chi
-{\rm arg Tr}{\cal U}(T)\rho(0)]
\end{eqnarray}
which are experimentally observable are manifestly gauge invariant. Here $\chi$ stands for the 
variable $U(1)$ phase (difference) in the interference beams.
We note that the gauge invariance of the interference pattern 
(2.39) does not hold  in the sense of the projective Hilbert space (2.25) in the conventional formulation~\cite{sjoqvist,singh}, which is 
related to the fact that the projective Hilbert space defined 
by (2.25) is not consistent with the superposition 
principle to describe interference~\cite{fujikawa3}.
 
\section{ Exactly solvable example}

We discuss the model
\begin{eqnarray}
&&\hat{H}=-\mu\hbar\vec{B}(t)\vec{\sigma},\nonumber\\
&&\vec{B}(t)=B(\sin\theta\cos\varphi(t), 
\sin\theta\sin\varphi(t),\cos\theta )
\end{eqnarray}
with $\varphi(t)=\omega t$ and constant $\omega,\ B$ and  
$\theta$. This model has been analyzed in the past by various 
authors by using the adiabatic approximation~\cite{berry}. It 
has been recently shown that this model is exactly treated in 
the 
framework of the second quantized formulation~\cite{fujikawa4,
fujikawa6}. 

The exact effective Hamiltonian (2.6) is given by 
\begin{eqnarray}
&&\hat{H}_{eff}(t)=[-\mu\hbar B
-\frac{(1+\cos\theta)}{2}\hbar\omega]\hat{b}^{\dagger}_{+}
\hat{b}_{+}\nonumber\\
&&+[\mu\hbar B-\frac{1-\cos\theta}{2}\hbar\omega]
\hat{b}^{\dagger}_{-}\hat{b}_{-}
-\frac{\sin\theta}{2}\hbar\omega
[\hat{b}^{\dagger}_{+}\hat{b}_{-}+
\hat{b}^{\dagger}_{-}\hat{b}_{+}]
\end{eqnarray}
if one uses the instantaneous eigenstates
\begin{eqnarray}
\hat{H}(t)v_{\pm}(t)=\mp\mu\hbar Bv_{\pm}(t)
\end{eqnarray}
as the complete basis set in (2.1) and the expansion
$\hat{\psi}(t)=\sum\hat{b}_{n}v_{n}(t)$.
This effective $H_{eff}$ is not diagonal, but it is 
{\em diagonalized} if one performs a unitary transformation
\begin{eqnarray}
\left(\begin{array}{c}
     \hat{b}_{+}(t)\\
     \hat{b}_{-}(t)
     \end{array}\right)
&=&
\left(\begin{array}{cc}
 \cos\frac{1}{2}\alpha&-\sin\frac{1}{2}\alpha\\
 \sin\frac{1}{2}\alpha &\cos\frac{1}{2}\alpha
            \end{array}\right)
\left(\begin{array}{c}
           \hat{c}_{+}(t)\\
           \hat{c}_{-}(t)
            \end{array}\right)
\end{eqnarray}
with a constant $\alpha$ satisfying the parameter equation 
\begin{eqnarray}
\tan\alpha=\frac{\hbar\omega\sin\theta}{2\mu\hbar B+\hbar\omega
\cos\theta}.
\end{eqnarray}
The corresponding new basis vectors are then explicitly given by 
\begin{eqnarray}
w_{+}(t)&=&\left(\begin{array}{c}
            \cos\frac{1}{2}(\theta-\alpha) e^{-i\varphi(t)}\\
            \sin\frac{1}{2}(\theta-\alpha)
            \end{array}\right),  
w_{-}(t)=\left(\begin{array}{c}
            \sin\frac{1}{2}(\theta-\alpha) e^{-i\varphi(t)}\\
            -\cos\frac{1}{2}(\theta-\alpha)
            \end{array}\right)
\end{eqnarray}
which satisfies $\hat{\psi}(t)=\sum\hat{b}_{n}v_{n}(t)=\sum
\hat{c}_{n}w_{n}(t)$.
These new basis vectors are periodic $w_{\pm}(0)=w_{\pm}(T)$ 
with $T=\frac{2\pi}{\omega}$,
and one can confirm 
\begin{eqnarray}
&&w_{\pm}^{\dagger}(t)\hat{H}w_{\pm}(t)
=\mp \mu\hbar B\cos\alpha\nonumber\\
&&w_{\pm}^{\dagger}(t)i\hbar\partial_{t}w_{\pm}(t)
=\frac{\hbar\omega}{2}(1\pm\cos(\theta-\alpha)).
\end{eqnarray}

The effective Hamiltonian $H_{eff}$ (3.2) is now diagonalized
in terms of $w_{\pm}(t)$, 
and thus  the {\em exact} solution of the Schr\"{o}dinger eq.,
$i\hbar\partial_{t}\psi(t)=\hat{H}\psi(t)$, is given by
\begin{eqnarray}
\psi_{\pm}(t)
&=&w_{\pm}(t)\exp\{-\frac{i}{\hbar}\int_{0}^{t}dt^{\prime}
[w_{\pm}^{\dagger}(t^{\prime})\hat{H}w_{\pm}(t^{\prime})
\nonumber\\
&&-w_{\pm}^{\dagger}(t^{\prime})i\hbar\partial_{t^{\prime}}
w_{\pm}(t^{\prime})]\}
\end{eqnarray}
if one uses the formula (2.8).
This amplitude may be regarded either as an exact version of the 
adiabatic phase or as a non-adiabatic cyclic phase in our 
formulation in (2.21).

We examine the two extreme limits of this formula:\\
(i) For the {\em adiabatic limit} 
$\hbar\omega/(\hbar\mu B)\ll 1$, the parameter equation (3.5)
gives 
\begin{eqnarray}
\alpha\simeq[\hbar\omega/2\hbar\mu B]\sin\theta, 
\end{eqnarray}
and if one sets 
$\alpha=0$ in the exact solution (3.8), one recovers the 
ordinary Berry phase~\cite{berry, simon}
\begin{eqnarray}
\psi_{\pm}(T)&\simeq&\exp\{i\pi(1\pm\cos\theta) \}\nonumber\\
&&\times\exp\{\pm\frac{i}{\hbar}\int_{0}^{T}dt
\mu\hbar B\}v_{\pm}(T)
\end{eqnarray}
where the first exponential factor stands for the 
"monopole-like phase" and
\begin{eqnarray}
v_{+}(t)&=&\left(\begin{array}{c}
            \cos\frac{1}{2}\theta e^{-i\varphi(t)}\\
            \sin\frac{1}{2}\theta
            \end{array}\right), 
v_{-}(t)=\left(\begin{array}{c}
            \sin\frac{1}{2}\theta e^{-i\varphi(t)}\\
            -\cos\frac{1}{2}\theta
            \end{array}\right).
\end{eqnarray}
(ii) For the {\em non-adiabatic} limit 
$\hbar\mu B/(\hbar\omega)\ll 1$, the parameter equation (3.5)
gives
\begin{eqnarray}
\theta-\alpha\simeq[2\hbar\mu B/\hbar\omega]\sin\theta
\end{eqnarray}
and if one sets $\alpha=\theta$ in the exact solution (3.8), one 
obtains the trivial phase
\begin{eqnarray}
\psi_{\pm}(T)
&\simeq&w_{\pm}(T)\exp\{\pm\frac{i}{\hbar}\int_{0}^{T}dt
[\mu\hbar B\cos\theta]\}
\end{eqnarray}
with
\begin{eqnarray}
w_{+}(t)&=&\left(\begin{array}{c}
            e^{-i\varphi(t)}\\
            0
            \end{array}\right), \nonumber\\  
w_{-}(t)&=&\left(\begin{array}{c}
            0\\
            -1
            \end{array}\right).
\end{eqnarray}
This shows that the ``monopole-like singularity'' is smoothly 
connected to a trivial phase in the exact solution, and thus 
the geometric phase is {\em topologically trivial}~\cite{fujikawa6}.

The adiabatic and non-adiabatic phases are treated in a unified 
manner in the present second quantized formulation, and thus 
this example shows that all the geometric phases are 
topologically trivial.

\section{ Chiral anomaly}

We consider the evolution operator 
\begin{eqnarray}
\int {\cal D}\bar{\psi}{\cal D}\psi \exp\{i\int d^{4}x
[\bar{\psi}i\gamma^{\mu}(\partial_{\mu}-igA_{\mu})\psi]\}
\end{eqnarray}
for the Dirac fermion $\psi(t,\vec{x})$ inside the background
gauge field $A_{\mu}(t,\vec{x})$.
The chiral anomaly in gauge field theory is understood in
path integrals as arising from the non-trivial Jacobian under 
the chiral transformation. 
For an infinitesimal chiral transformation of field variables
\begin{eqnarray}
\psi(x)\rightarrow e^{i\omega(x)\gamma_{5}}\psi(x),\ \ \ \ 
\bar{\psi}(x)\rightarrow \bar{\psi}(x)e^{i\omega(x)\gamma_{5}}
\end{eqnarray}
we have a non-trivial Jacobian
\begin{eqnarray}
{\cal D}\bar{\psi}{\cal D}\psi\rightarrow 
\exp\{-i\int d^{4}x \omega(x)\frac{g^{2}}{16\pi^{2}}
\epsilon^{\mu\nu\alpha\beta}F_{\mu\nu}F_{\alpha\beta}\}{\cal D}\bar{\psi}{\cal D}\psi
\end{eqnarray}
which is valid for a general class of regularization including 
the lattice gauge theory~\cite{fujikawa-suzuki}. The Jacobian
factor is identified with the chiral anomaly, and the 
integrated or summed Jacobian  is called the Wess-Zumino 
term~\cite{wess}.

Some of the known essential and general properties of the 
quantum anomalies are~\cite{fujikawa-suzuki}:\\
1. The anomalies are not recognized by a naive manipulation of 
the classical Lagrangian or action (or by a naive canonical 
manipulation in operator formulation), which leads to the naive 
N\"{o}ther's theorem.\\
2. The quantum anomaly is related to the quantum breaking of 
classical symmetries (and the failure of the naive N\"{o}ther's 
theorem). For example, the Gauss law operator (or BRST charge)
becomes time-dependent and thus it cannot be used to specify 
physical states in anomalous gauge theory.\\
3. The quantum anomalies are generally associated with an 
infinite number of degrees of freedom. The anomalies in the 
practical calculation are thus 
closely related to the regularization, though the anomalies by 
themselves are perfectly finite.\\
4. In the path integral formulation, the anomalies are 
recognized as non-trivial Jacobians for the change of 
path integral variables associated with classical symmetries,
as is explained above.
\\

None of these essential properties are shared with the 
geometric phases discussed in Sections 2 and 3.
One rather recognizes the following basic differences between
the geometric phases and chiral anomaly~\cite{fujikawa5}:\\
1.The Wess-Zumino term, which is obtained by a sum of the  
 infinitesimal Jacobian such as in (4.3), 
 is added to the classical action in path integrals, whereas 
the geometric term appears {\em inside} the classical action 
sandwiched by field variables as in (2.6) 
\begin{eqnarray}
\hat{H}_{eff}(t)&=&\sum_{n,m}\hat{b}_{n}^{\dagger}(t)[
\int d^{3}x v_{n}^{\star}(t,\vec{x})\hat{H}(t)v_{m}(t,\vec{x})
\nonumber\\
&&-\int d^{3}x v_{n}^{\star}(t,\vec{x})
i\hbar\frac{\partial}{\partial t}v_{m}(t,\vec{x})]
\hat{b}_{m}(t).
\end{eqnarray}
The geometric phase thus depends on each state in the Fock space generated by $\hat{b}_{n}^{\dagger}$, whereas the chiral anomaly is state-independent.
\\
2. The topology of chiral anomaly, which is provided by given 
gauge field,
is exact, whereas the topology of the adiabatic geometric phase,
which is valid only approximately in the adiabatic limit, is 
trivial as we have shown in Section 3.\\
3. The geometric phases are basically different from the 
topologically exact objects such as the Aharonov-Bohm phase or 
chiral anomaly. For example, the Aharonov-Bohm phase is 
identical for adiabatic or non-adiabatic motion of the electron.
\\
4. {\em Similarity} between the geometric phase and a special 
class of chiral anomaly was noted by M. Stone on the basis of a 
model~\cite{stone}
\begin{eqnarray}
{\cal H}(t)=\frac{\vec{L}^{2}}{2I}-\psi^{\dagger}\mu{\bf n}(t)
\cdot\vec{\sigma}\psi
\end{eqnarray}
where ${\bf n}(t)$ plays a role of the magnetic field in (3.1)
which acts on the spin $\vec{\sigma}$, and $\vec{L}$ induces 
the rotation of ${\bf n}(t)$.
But it is obvious from our analysis of topological properties 
in Section 3 that these two notions are fundamentally different.
\\
5. The topology of Berry's phase is valid only when the 
adiabatic approximation is strictly valid, whereas the anomaly 
appears in field theory only when the adiabatic approximation 
{\em fails} in a 
version of the Hamiltonian analysis~\cite{nelson}. Thus these 
notions cannot be compatible.

\section{Conclusion}

We have illustrate the advantages of the second quantized 
formulation of all the geometric phases. The second quantized
formulation is located in between the first quantization and 
field theory, and thus it is convenient to compare the 
geometric phase with other phases such as chiral anomaly. 
We clarified the basic differences between these two notions.

In the early literature on the geometric phase, the similarity 
between the geometric phase and other phases such as the chiral
anomaly and the Aharonov-Bohm phase, was often emphasized.  But
in view of the wide use of the loosely defined  
terminology ``geometric phase'' in various fields in physics 
today, it is our opinion that a more precise 
distinction of ``identical phenomena'' from ``similar 
phenomena'' is important. To be precise, what we are suggesting 
is to call chiral anomaly as chiral anomaly,
Wess-Zumino term as Wess-Zumino term, and Aharovov-Bohm phase 
as Aharonov-Bohm
phase, etc., since those terminologies convey  very clear 
messages and well-defined physical contents which the majority 
in physics community can readily recognize. Even in this sharp 
definition of terminology, one should still be able to clearly 
identify the geometric phase and its physical 
characteristics, which  are intrinsic to the geometric 
phase and cannot be described by other notions.

\end{document}